\documentclass[5p,fleqn,twocolumn,final]{elsarticle}
\biboptions{sort,compress}
\usepackage{showkeys}
\usepackage{amssymb}
\usepackage[mathscr]{eucal}
\usepackage{amsmath}
\usepackage{graphicx}
\usepackage{pstricks}
\usepackage{wrapfig}
\usepackage{amsbsy}

\journal{Journal of Magnetism and Magnetic Materials}

\begin{document}

\begin{frontmatter}

\title{Spin and charge fluctuations in the Hubbard model}

\author{A. Sherman}
\ead{alekseis@ut.ee}


\address{Institute of Physics, University of Tartu, W. Ostwaldi Str 1, 50411 Tartu, Estonia}

\begin{abstract}
Using the strong coupling diagram technique for calculating the electron Green's function of the two-dimensional Hubbard model we have summed infinite sequences of ladder diagrams, which describe interactions of electrons with spin and charge fluctuations. For sufficiently low temperatures and doping a pronounced four-band structure is observed in  spectral functions. Its appearance is related to the proximity of the transition to the long-range antiferromagnetic order.
\end{abstract}


\end{frontmatter}

\section{Introduction}
The strong coupling diagram technique \cite{Vladimir,Metzner,Craco,Pairault,Sherman06} is an approach for investigating systems with moderate to strong electron correlations. This method uses the expansion in powers of the kinetic energy operator on the assumption that it is smaller than the potential part of a Hamiltonian. In application to the repulsive Hubbard model this approach allowed one to describe the Mott metal-insulator transition. In a local approximation for the irreducible part the method reproduces electron spectral functions in reasonable agreement with Monte Carlo data for high temperatures in a wide range of doping \cite{Sherman15}.

In this work we improve the local approximation by including into consideration an infinite sequences of diagrams, which describe interactions of electrons with spin and charge fluctuations. As a consequence the irreducible part acquires a momentum dependence and the four-band structure appears in electron spectral functions at sufficiently low temperatures and doping. This structure stems from a short-range antiferromagnetic order with a large enough correlation length.

\section{Main formulas}
The Hamiltonian of the two-dimensional (2D) fermionic Hubbard model \cite{Hubbard63} reads
\begin{equation}\label{Hamiltonian}
H=\sum_{\bf ll'\sigma}t_{\bf ll'}a^\dagger_{\bf l'\sigma}a_{\bf l\sigma}
+\frac{U}{2}\sum_{\bf l\sigma}n_{\bf l\sigma}n_{\bf l,-\sigma},
\end{equation}
where 2D vectors ${\bf l}$ and ${\bf l'}$ label sites of a square plane lattice, $\sigma=\pm 1$ is the spin projection, $a^\dagger_{\bf l\sigma}$ and $a_{\bf l\sigma}$ are electron creation and annihilation operators, $t_{\bf ll'}$ is the hopping constant, $U$ is the on-site Coulomb repulsion and $n_{\bf l\sigma}=a^\dagger_{\bf l\sigma}a_{\bf l\sigma}$. In this work only the nearest neighbor hopping constant $t$ is supposed to be nonzero.

We shall consider the electron Green's function
\begin{equation}\label{Green}
G({\bf l'\tau',l\tau})=\langle{\cal T}\bar{a}_{\bf l'\sigma}(\tau')
a_{\bf l\sigma}(\tau)\rangle,
\end{equation}
where the statistical averaging denoted by the angular brackets and time dependencies are determined by the operator ${\cal H}=H-\mu\sum_{\bf l\sigma}n_{\bf l\sigma}$ with the chemical potential $\mu$. ${\cal T}$ is the time-or\-de\-r\-ing operator which arranges operators from right to left in ascending order of times $\tau$. To consider the case of strong electron correlations, $U\gg t$, for calculating this function we use the strong coupling diagram technique \cite{Vladimir,Metzner,Craco,Pairault,Sherman06}. As the Coulomb repulsion in the unperturbed Hamiltonian $H_0$ is a quartic function of electron operators, Wick's theorem cannot be used for disentangling averages in the power expansion. However, $H_0$ is the sum of local terms, which allows one to represent any of these averages as a sum of all possible local cumulants of the operators $a^\dagger_{\bf l\sigma}$ and $a_{\bf l\sigma}$. As a result terms of the series expansion become products of on-site cumulants and hopping constants. If we present graphically cumulants as circles, each such a term will look like a sequence of circles and connecting them arrowed lines corresponding to $t_{\bf ll'}$.

\begin{figure}[t]
\centerline{\resizebox{0.9\columnwidth}{!}{\includegraphics{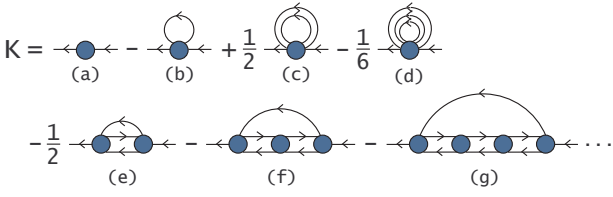}}}
\caption{Diagrams of several lowest orders in $K({\bf k}j)$.} \label{Fig1}
\end{figure}

As in the weak-coupling perturbation theory, the linked-cluster theorem allows one to discard disconnected diagrams and to carry out partial summations in connected diagrams. The diagram is said to be irreducible if it cannot be divided into two disconnected parts by cutting some hopping line. The sum of all irreducible diagrams is termed the irreducible part $K({\bf l'\tau',l\tau})$. In terms of this quantity the equation for the Green's function reads
\begin{equation}\label{Larkin}
G({\bf k}j)=\left\{\left[K({\bf k}j)\right]^{-1}-t_{\bf k}\right\}^{-1},
\end{equation}
where ${\bf k}$ is the 2D wave vector and $j$ is an integer in the Matsubara frequency $\omega_j=(2j+1)\pi T$ with the temperature $T$. Diagrams of several lowest orders in $K({\bf k}j)$ are shown in Fig.~\ref{Fig1} with their signs and prefactors.
Thanks to the possibility of partial summation, bare internal lines $t_{\bf k}$ in Fig.~\ref{Fig1} can be transformed into dressed ones,
\begin{equation}\label{hopping}
\theta({\bf k}j)=t_{\bf k}+t^2_{\bf k}G({\bf k}j).
\end{equation}
As was shown in \cite{Sherman15}, $K({\bf k}j)$ approximated by the sum of diagrams (a) and (b) allows one to describe the Mott transition and to obtain spectral functions in reasonable agreement with Monte Carlo results for $T\gtrsim t$. In this approximation the irreducible part is independent of ${\bf k}$.

\begin{figure}[t]
\centerline{\resizebox{0.9\columnwidth}{!}{\includegraphics{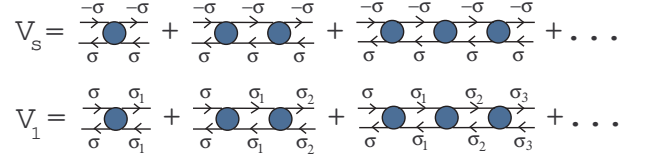}}}
\caption{Two types of infinite sums of ladders contributing to $K({\bf k}j)$.} \label{Fig2}
\end{figure}

In this work, in addition to diagrams (a) and (b) we take into account an infinite sequence of diagrams containing ladder inserts. Several diagrams of this type are shown in the second row of Fig.~\ref{Fig1}. These diagrams are of interest, since sums of ladders represent charge and spin susceptibilities \cite{Sherman07}. Hence the diagrams describe interactions of electrons with spin and charge fluctuations. They introduce momentum dependence in the irreducible part,
\begin{equation}\label{irreducible}
K({\bf k}j)=C_1(j)-\frac{T}{N}\sum_{{\bf k'}j'\sigma'}\theta({\bf k-k'},j')V_{\bf k'}(j\sigma,j\sigma,j'\sigma',j'\sigma'),
\end{equation}
where $C_1$ is the first-order cumulant in the diagram (a)
\begin{equation}\label{C_1}
C_1(j)=Z^{-1}\big[\big(e^{\mu/T}+1\big)g_{01}(j)
+\big(e^{(2\mu-U)/T}+e^{\mu/T}\big)g_{12}(j)\big],
\end{equation}
$Z=1+2\exp(\mu/T)+\exp[(2\mu-U)/T]$, $g_{01}(j)=(i\omega_j+\mu)^{-1}$, $g_{12}(j)=(i\omega_j+\mu-U)^{-1}$, $N$ is the number of sites and $V_{\bf k'}(j\sigma,j\sigma,j'\sigma',j'\sigma')$ is the infinite sum of ladder diagrams of the type shown in Fig.~\ref{Fig2}.
Four frequencies and spins in $V$ are variables of external legs of diagrams and ${\bf k'}$ is the moment transferred through them. To simplify the problem only second-order cumulants were used in ladders.

The quantity $V$ satisfies the Bethe-Salpeter equation (BSE). Cumulants retain spin projections. Consequently, quantities $V$ in Eq.~(\ref{irreducible}) with $\sigma'=-\sigma$ and $\sigma'=\sigma$ are calculated from two different BSE corresponding to the series in Fig.~\ref{Fig2}. One of them is denoted by $V_s$. The subscript points to the analogy with ladder diagrams in the spin susceptibility \cite{Sherman07}.

Equations describing second-order cumulants in BSE are given in \cite{Sherman06,Sherman07}. They are rather cumbersome. However, the equations can be significantly simplified in the case $T\ll\mu\ll U-T$.
For $U\gg T$ this range contains most interesting cases from half-filling ($\mu=U/2$) to moderate doping. For such $\mu$ the second-order cumulants read
\begin{eqnarray}
&&C_2(j+\nu,\sigma;j\sigma;j',-\sigma;j'+\nu,-\sigma)=-\bigg(\frac{\delta_{jj'}}{2T}+ \frac{\delta_{\nu 0}}{4T}\bigg)\nonumber\\
&&\quad\times a_1(j'+\nu)a_1(j)+C_2'(jj'\nu),\label{C_2s}\\
&&C_2(j+\nu,\sigma;j\sigma';j'\sigma';j'+\nu,\sigma)=\frac{1}{4T}\big[\delta_{jj'}\big(1-2 \delta_{\sigma\sigma'}\big)\nonumber\\
&&\quad+\delta_{\nu 0}\big(2-\delta_{\sigma\sigma'}\big)\big]a_1(j'+\nu)a_1(j)\nonumber\\
&&\quad-\delta_{\sigma,-\sigma'}C_2'(jj'\nu), \label{C_2c}
\end{eqnarray}
where
\begin{eqnarray*}
&&C_2'(jj'\nu)=\frac{1}{2}\big[a_1(j'+\nu)a_2(jj')\nonumber\\
&&\quad+a_4(j'+\nu,j+\nu)a_3(jj')+a_2(j'+\nu,j+\nu)a_1(j)\nonumber\\
&&\quad+a_3(j'+\nu,j+\nu)a_4(jj')\big], \label{terms}\\
&&a_1(j)=g_{01}(j)-g_{12}(j),\quad a_2(jj')=g_{01}(j)g_{01}(j'),\nonumber\\
&&a_3(jj')=g_{12}(j)-g_{01}(j'),\quad a_4(jj')=a_1(j)g_{12}(j').\nonumber
\end{eqnarray*}
Expression (\ref{C_2s}) is used in the BSE for $V_s$,
\begin{eqnarray}\label{BS}
&&V_{s{\bf k}}(j+\nu,\sigma;j\sigma;j',-\sigma;j'+\nu,-\sigma)\nonumber\\
&&\quad=C_2(j+\nu,\sigma;j\sigma;j',-\sigma;j'+\nu,-\sigma)\nonumber\\
&&\quad+T\sum_{\nu'}C_2(j+\nu,\sigma;j+\nu',\sigma;j'+\nu',-\sigma;j'+\nu,-\sigma)\nonumber\\
&&\quad\times{\cal T}_{\bf k}(j+\nu',j'+\nu')\nonumber\\
&&\quad\times V_{s{\bf k}}(j+\nu',\sigma;j\sigma;j',-\sigma;j'+\nu',-\sigma),
\end{eqnarray}
while Eq.~(\ref{C_2c}) is employed in an equation for $V_1$. In Eq.~(\ref{BS})
\begin{equation}\label{calt}
{\cal T}_{\bf k}(jj')=\frac{1}{N}\sum_{\bf k'}\theta({\bf k+k'},j)\theta({\bf k'}j').
\end{equation}

Substituting Eq.~(\ref{C_2s}) into (\ref{BS}) we get
\begin{eqnarray}\label{Vsk}
&&V_{s{\bf k}}(j+\nu,j,j',j'+\nu)=\frac{f_1({\bf k},j+\nu,j'+\nu)}{2}\nonumber\\
&&\quad\times\bigg\{2C_2(j+\nu,j,j',j'+\nu)+\bigg[a_2(j'+\nu,j+\nu)\nonumber\\
&&\quad-\frac{\delta_{jj'}}{T}a_1(j'+\nu)\bigg]y_1({\bf k}jj')+a_1(j'+\nu)y_2({\bf k}jj')\nonumber\\
&&\quad+a_4(j'+\nu,j+\nu)y_3({\bf k}jj')\nonumber\\
&&\quad+a_3(j'+\nu,j+\nu)y_4({\bf k}jj')\bigg\},
\end{eqnarray}
where for brevity we dropped spin indices,
\begin{equation}
f_1({\bf k}jj')=\bigg[1+\frac{1}{4}a_1(j)a_1(j'){\cal T}_{\bf k}(jj')\bigg]^{-1}\label{f1},
\end{equation}
\begin{eqnarray}
&&y_i({\bf k}jj')=T\sum_\nu a_i(j+\nu,j'+\nu){\cal T}_{\bf k}(j+\nu,j'+\nu)\nonumber\\
&&\quad\times V_{s{\bf k}}(j+\nu,j,j',j'+\nu) \label{y_i}.
\end{eqnarray}
Equations for $y_i({\bf k}jj')$ are derived from Eq.~(\ref{Vsk}),
\begin{eqnarray}\label{eq_for_y}
&&y_i({\bf k}jj')=b_i({\bf k}jj')+\bigg[c_{i2}({\bf k}jj')-\frac{\delta_{jj'}}{T}c_{i1}({\bf k}jj')\bigg] \nonumber\\
&&\quad\times y_1({\bf k}jj')+c_{i1}({\bf k}jj')y_2({\bf k}jj')+c_{i4}({\bf k}jj')y_3({\bf k}jj')\nonumber\\
&&\quad+c_{i3}({\bf k}jj')y_4({\bf k}jj'),
\end{eqnarray}
where
\begin{eqnarray}
&&b_i({\bf k}jj')=-\frac{1}{4}a_i(jj')a_1(j)a_1(j'){\cal T}_{\bf k}(jj')f_1({\bf k}jj')\nonumber\\
&&\quad+\bigg[a_2(jj')-\frac{\delta_{jj'}}{T}a_1(j)\bigg]c_{i1}({\bf k}jj')+a_1(j)c_{i2}({\bf k}jj')\nonumber\\
&&\quad+a_4(jj')c_{i3}({\bf k}jj')+a_3(jj')c_{i4}({\bf k}jj'),\label{b_i}\\
&&c_{ii'}({\bf k}jj')=\frac{T}{2}\sum_\nu a_i(j+\nu,j'+\nu)a_{i'}(j'+\nu,j+\nu)\nonumber\\
&&\quad\times{\cal T}_{\bf k}(j+\nu,j'+\nu)f_1({\bf k},j+\nu,j'+\nu).\label{c_ii}
\end{eqnarray}
Notice that $c_{ii'}({\bf k}jj')$ depend only on the difference $j-j'$. Thus, the solution of BSE (\ref{BS}) was reduced to the system of four linear equations (\ref{eq_for_y}) with respect to $y_i({\bf k}jj')$, $i=1-4$. The above equations are similar to those used in \cite{Sherman07} for investigating the spin susceptibility.

To solve the BSE for $V_1$ we divide it into two parts,
\begin{eqnarray}
&&V_{c\bf k}(j+\nu,j,j',j'+\nu)\nonumber\\
&&\quad=\sum_{\sigma'}V_{1{\bf k}}(j+\nu,\sigma';j\sigma;j'\sigma;j'+\nu,\sigma'), \nonumber\\[-1ex]
&&\label{Vsc}\\[-1ex]
&&V^-_{\bf k}(j+\nu,j,j',j'+\nu)\nonumber\\
&&\quad=\sigma\sum_{\sigma'}\sigma' V_{1{\bf k}}(j+\nu,\sigma';j\sigma;j'\sigma;j'+\nu,\sigma').\nonumber
\end{eqnarray}
Equations for $V_{\bf k}^-$ are identical to Eqs.~(\ref{Vsk})--(\ref{c_ii}) and, consequently, $V_{\bf k}^-=V_{s\bf k}$. The equation for $V_{c\bf k}$ reads
\begin{eqnarray}\label{Vck}
&&V_{c\bf k}(j+\nu,j,j',j'+\nu)=\frac{f_2({\bf k},j+\nu,j'+\nu)}{2}\nonumber\\
&&\quad\times\big[2C_2^+(j+\nu,j,j',j'+\nu)-a_2(j'+\nu,j+\nu)\nonumber\\
&&\quad\times z_1({\bf k}jj')-a_1(j'+\nu)z_2({\bf k}jj')\nonumber\\
&&\quad-a_4(j'+\nu,j+\nu)z_3({\bf k}jj')\nonumber\\
&&\quad-a_3(j'+\nu,j+\nu)z_4({\bf k}jj')\big],
\end{eqnarray}
where
\begin{eqnarray}
&&f_2({\bf k}jj')=\bigg[1-\frac{3}{4}a_1(j)a_1(j'){\cal T}_{\bf k}(jj')\bigg]^{-1},\label{f2}\\
&&C^+_2(j+\nu,j,j',j'+\nu)\nonumber\\
&&\quad=\sum_{\sigma'}C_2(j+\nu,\sigma';j\sigma;j'\sigma;j'+\nu,\sigma'), \label{C+2}\\
&&z_i({\bf k}jj')=T\sum_\nu a_i(j+\nu,j'+\nu){\cal T}_{\bf k}(j+\nu,j'+\nu)\nonumber\\
&&\quad\times V_{c\bf k}(j+\nu,j,j',j'+\nu) \label{z_i}.
\end{eqnarray}
Cumulant (\ref{C_2c}) is used for calculating $C^+_2$ in Eq.~(\ref{C+2}). Quantities $z_i({\bf k}jj')$ are obtained from the four linear equations
\begin{eqnarray}\label{eq_for_z}
&&z_i({\bf k}jj')=d_i({\bf k}jj')-e_{i2}({\bf k}jj')z_1({\bf k}jj')\nonumber\\
&&\quad-e_{i1}({\bf k}jj')z_2({\bf k}jj')-e_{i4}({\bf k}jj')z_3({\bf k}jj')\nonumber\\
&&\quad-e_{i3}({\bf k}jj')z_4({\bf k}jj')
\end{eqnarray}
with
\begin{eqnarray}
&&d_i({\bf k}jj')=\frac{3}{4}a_i(jj')a_1(j)a_1(j'){\cal T}_{\bf k}(jj')f_2({\bf k}jj')\nonumber\\
&&\quad-a_2(jj')e_{i1}({\bf k}jj')-a_1(j)e_{i2}({\bf k}jj')\nonumber\\
&&\quad-a_4(jj')e_{i3}({\bf k}jj')-a_3(jj')e_{i4}({\bf k}jj'),\label{d_i}\\
&&e_{ii'}({\bf k}jj')=\frac{T}{2}\sum_\nu a_i(j+\nu,j'+\nu)a_{i'}(j'+\nu,j+\nu)\nonumber\\
&&\quad\times{\cal T}_{\bf k}(j+\nu,j'+\nu)f_2({\bf k},j+\nu,j'+\nu).\label{d_ii}
\end{eqnarray}
The quantities $e_{ii'}({\bf k}jj')$ depend only on the difference $j-j'$. The above equations for $V_{c\bf k}$ are similar to those derived in \cite{Sherman07} for the charge susceptibility.

Finally, the irreducible part (\ref{irreducible}) reads
\begin{eqnarray}\label{irreducible2}
&&K({\bf k}j)=C_1(j)-\frac{T}{N}\sum_{{\bf k'}j'}\theta({\bf k-k'},j')\bigg[\frac{3}{2}V_{s\bf k'}(jjj'j')\nonumber\\
&&\quad+\frac{1}{2}V_{c\bf k'}(jjj'j')\bigg].
\end{eqnarray}

\section{Results and discussion}
Equations (\ref{Larkin}), (\ref{Vsk}), (\ref{eq_for_y}), (\ref{Vck}), (\ref{eq_for_z}) and (\ref{irreducible2}) were self-consistently solved by iteration, starting with Green's function of the Hubbard-I approximation \cite{Hubbard63}. It is obtained from Eq.~(\ref{Larkin}) if the irreducible part is approximated by the first-order cumulant (\ref{C_1}) \cite{Vladimir}. In this work we use a 8$\times$8 lattice. The analytic continuation to real frequencies was carried out using the continued-fraction method from \cite{Vidberg}. In some cases in this procedure some poles of the obtained retarded Green's function got to the upper complex half-plane, though being located very close to the real axis. To avoid negative values in spectral functions we set $A({\bf k}\omega)=|{\rm\, Im}G({\bf k}\omega)|/\pi$. The below figures were calculated for $U=8t$ and $T\approx 0.67t$. Notice that the equations of the previous section describe the establishment of the long-range antiferromagnetic order at half-filling for $T_{\rm AF}\approx 0.4t$. This fact is connected with vanishing the denominator of the system (\ref{eq_for_y}). The transition temperature is nonzero pointing to the violation of the Mermin-Wagner theorem \cite{Mermin}. Partly this is connected with the finiteness of the lattice. The temperature used for the calculations was chosen so that to obtain the short-range order with the correlation length equalling to a few lattice spacings. In the below figures $t$ and the lattice spacing are set as units of energy and length, respectively.
\begin{figure}[t]
\centerline{\resizebox{0.7\columnwidth}{!}{\includegraphics{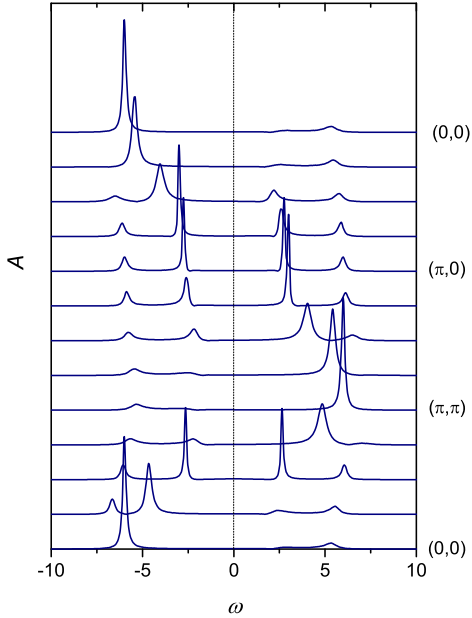}}}
\caption{The spectral function $A({\bf k}\omega)$ calculated in a 8$\times$8 lattice for $U=8$, $n=1$, $T\approx 0.67$ and wave vectors along the symmetry lines of the Brillouin zone. The dashed line represents the Fermi energy.} \label{Fig3}
\end{figure}

\begin{figure}[t]
\centerline{\resizebox{0.7\columnwidth}{!}{\includegraphics{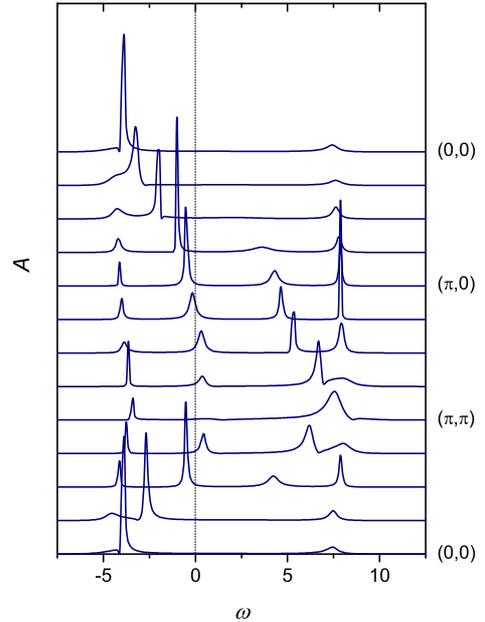}}}
\caption{Same as in Fig.~\protect\ref{Fig3}, but for $n\approx 0.9$.} \label{Fig4}
\end{figure}
The obtained spectral functions are shown in Fig.~\ref{Fig3} for the case of half-filling, $n=1$, and in Fig.~\ref{Fig4} for $n\approx 0.9$. Here $n=2N^{-1}\sum_{\bf k}\int_{-\infty}^\infty[\exp(\omega/T)+1]^{-1}A({\bf k}\omega)d\omega$ is the electron concentration. A pronounced four-band structure is seen in the spectra. Analogous spectral functions containing four intensive maxima in the major part of the Brillouin zone were earlier obtained in Monte Carlo simulations \cite{Preuss,Grober} and in different analytical approaches \cite{Senechal,Ovchinnikov,Dahnken,Kyung}. As follows from Eq.~(\ref{irreducible2}), the interaction with spin and charge fluctuations mixes electron states with momenta, which differ by wave vectors of sharp peaks in $V_{s\bf k}$ and $V_{c\bf k}$. In the considered case the system is near the antiferromagnetic transition, and $V_{s\bf k}$ is sharply peaked at the antiferromagnetic momentum ${\bf Q}=(\pi,\pi)$.
The four-band structure is related to this peak, as follows from above equations and from calculations -- though the structure is well seen for small doping in Fig.~\ref{Fig4}, for $n\approx 0.8$ (not shown here) and higher $T$, when the peak in $V_{s\bf k}$ is degraded, spectral maxima are blurred and the structure disappears.

\section{Concluding remarks}
In this work, the strong coupling diagram technique was used for calculating the electron Green's function of the two-dimensional repulsive Hubbard model. In contrast to previous works in this approach, an infinite sequence of diagrams with ladder inserts was taken into account in the irreducible part. These diagrams describe interactions of electrons with spin and charge fluctuations. Obtained equations were self-consistently solved by iteration in a 8$\times$8 lattice for the Hubbard repulsion $U=8t$, $t$ being the nearest-neighbor hopping constant, and for the temperature $T\approx 0.67t$. The calculated spectral functions demonstrate the four-band structure, which is similar to those observed in results of some Monte Carlo simulations and cluster approaches. The structure is related to the proximity of the transition to the long-range antiferromagnetic ordering.

\section*{Acknowledgements}
This work was supported by the project IUT2-27 and the Estonian Scientific Foundation (grant ETF9371).

\end{document}